\documentclass[runningheads]{llncs}
\usepackage[table,xcdraw]{xcolor}
\usepackage{graphicx}
\usepackage{cite}
\usepackage{adjustbox}
\usepackage{amsmath}
\usepackage[english]{babel}
\usepackage{float}
\usepackage{hyperref}
\usepackage[all]{hypcap}
\usepackage{cancel}

\usepackage{svg}
\usepackage{tabularx}
\usepackage{multirow}
\newcolumntype{Y}{>{\centering\arraybackslash}X}
\usepackage{booktabs}
\usepackage{relsize}
\newcommand{\link}[1]{%
    {%
    \def\textcolor##1##2{##2}%
    \def\textbf##1{##1}%
    \def\textit##1{##1}%
    \edef\&{\string&}%
    \edef\_{\string_}%
    \xdef\tmp{\noexpand\href{#1}}}%
 \tmp{#1}%
}

\hypersetup{linktoc=all}

\AtBeginDocument{}

\begin{document}
\title{Echo from noise: synthetic ultrasound image generation using diffusion models for real image segmentation}
\titlerunning{Echo from noise: Synthetic ultrasound generation}
\author{
David Stojanovski \inst{1}\orcidID{0000-0002-7912-5249}\thanks{
This work was supported by the Wellcome/EPSRC Centre for Medical Engineering [WT203148/Z/16/Z] and by the National Institute for Health Research (NIHR) Biomedical Research Centre at Guy's and St Thomas' NHS Foundation Trust and King's College London. The views expressed are those of the author(s) and not necessarily those of the NHS, the NIHR or the Department of Health.
} \and
Uxio Hermida \inst{1}\orcidID{0000-0003-1671-3489} \and
Pablo Lamata \inst{1}\orcidID{0000-0002-3097-4928} \and
Arian Beqiri \inst{1,2}\orcidID{0000-0002-9169-1788} \and
Alberto Gomez \inst{1,2}\orcidID{0000-0002-7897-7589}}
\authorrunning{Stojanovski et al.}

\institute{
 King’s College London, School of Biomedical Engineering \& Imaging Sciences,
London, SE1 7EU, UK \email{\{first.last\}@kcl.ac.uk}\\ \and
 Ultromics Ltd., Oxford, OX4 2SU, UK \email{\{first.last\}@ultromics.com}\\
 david.stojanovski@kcl.ac.uk}

\maketitle

\begin{abstract}
We propose a novel pipeline for the generation of synthetic ultrasound images via Denoising Diffusion Probabilistic Models (DDPMs) guided by cardiac semantic label maps. We show that these synthetic images can serve as a viable substitute for real data in the training of deep-learning models for ultrasound image analysis tasks such as cardiac segmentation. To demonstrate the effectiveness of this approach, we generated synthetic 2D echocardiograms and trained a neural network for segmenting the left ventricle and left atrium. The performance of the network trained on exclusively synthetic images was evaluated on an unseen dataset of real images and yielded mean Dice scores of  88.6 $\pm 4.91$ , 91.9 $\pm 4.22$, 85.2 $\pm 4.83$ \% for left ventricular endocardium, epicardium and left atrial segmentation respectively. This represents a relative increase of $9.2$, $3.3$ and $13.9$ \% in Dice scores compared to the previous state-of-the-art. The proposed pipeline has potential for application to a wide range of other tasks across various medical imaging modalities.

\end{abstract}

\section{Introduction}
\subsection{Background and motivation}
Echocardiography (echo) is the most widely used method for evaluating the heart, because it is more cost-effective and safer than other imaging modalities, while providing high resolution images in real-time. However, a major drawback of echo is its heavy dependence on operators' expertise to obtain high-quality images and associated anatomical and functional measurements. 

Convolutional neural networks (CNNs) have shown great potential for automating medical image analysis tasks and are capable of accurately learning complex relevant features from large sets of data. A major challenge in the use of CNNs for medical imaging tasks is the need for labelling such large sets of data for model training. Further, CNN accuracy can be limited by the quality of the labels used during training, particularly in the presence of noise or other artefacts that can lead to inter-observer errors. Experienced cardiologists have been shown to have inter-observer errors up to 22\% when labelling common measurements in echo \cite{Armstrong2015}.

Previous research has shown the feasibility of generating realistic natural and medical images \cite{Zhu2017, Gilbert2021, Armanious2019}. Until recently, the state-of-the-art (SOTA) results were achieved with Generative Adversarial Networks (GANs) and CycleGANs. GANs are known to be notoriously difficult to train, due to training instability as the generator and discriminator are trained simultaneously, and the loss function can be highly non-convex, making it challenging to find the global minimum. Additionally, GANs can suffer from vanishing gradients, which can lead to slow or non-existent convergence. This often leads to failed training runs and great difficulty in reproducing results. GANs are also prone to mode collapse, where the generator learns to produce a limited set of outputs, which can be repeated instead of generating diverse samples \cite{pmlr-v139-feng21c}. This can happen when the discriminator is too strong in comparison to the generator, and rejects any samples that do not match the training data, forcing the generator to produce only a limited set of outputs. When using a guide image for synthetic image synthesis (e.g., to generate a synthetic medical image guided by an anatomical representation), CycleGANs have been the go-to technique of choice, but typically fail to reproduce the anatomy under large transformations of the guide images and collapse to anatomy seen in the training set, as illustrated later in this paper.

Denoising Diffusion Probabilistic models (DDPMs) are more recent generative models that are far less susceptible to the pitfalls of GAN based methods \cite{https://doi.org/10.48550/arxiv.2212.07501}. However, limited research has been performed on medical images, with a few notable examples \cite{Pinaya2022,Lyu2022}. No prior research studies have explored their use in semantically guided image synthesis, or to ultrasound imaging at all. In this paper, we use DDPMs to generate synthetic echo images and train a segmentation model. 

\subsection{Related works} 
The two primary Deep Learning (DL) techniques for generating synthetic images are GANs and DDPMs.

\subsubsection{Generative Adversarial Networks:}
GANs and the CycleGAN subclass have shown success in generating synthetic medical images. Examples include Cycle-MedGAN \cite{Armanious2019} that achieved $\>0.91$ Structural Similarity Index Measure (SSIM) in a Positron Emission Tomography to Computed Tomography (CT) unsupervised translation task. The SOTA in segmentation is presented by Gilbert et al. who used a CycleGAN architecture to generate images for training a network to segment an unseen real cardiac ultrasound test set. They achieved 79.4, 88.6 and 71.3 \% mean Dice scores on the left ventricle endocardium, left ventricle epicardium and left atrium respectively \cite{Gilbert2021b}.

\subsubsection{Diffusion Models:}
 Sohl-Dickstein et al. initially proposed DDPMs \cite{Sohl2015}, which have been used successfully for unsupervised image-to-image translation in both the natural and medical image domains. Pinaya et al. achieved a Fréchet inception distance (FID) of $7.8 \times 10^{-3}$ for brain Magnetic Resonance Imaging (MRI) generation versus an FID of $5 \times 10^{-4}$ for real brain MRI \cite{Pinaya2022}. Lyu et al. achieved $>0.85$ SSIM score for their conversion of CT to MRI images \cite{Lyu2022}. 

These works relate to the problem of image-to-image domain adaption. Instead, we use diffusion models to address the issue of limited labelled training data in echo image segmentation, by using a semantically guided network that receives an anatomical semantic label map to generate conditional synthetic echo images which adhere to the anatomy. Our work aims to combine the benefits of DDPMs with semantically guided medical imaging to synthetically increase dataset size and enable the creation of out-of-distribution images.

\subsection{Contributions of this study}
This is the first work to utilize DDPMs for generating medical images using semantic label maps as a source image for conditioning the generated image. To summarize our contributions, we: 1) Demonstrate that a semantic label map guided diffusion model can be trained to synthesize cardiac ultrasound images and matching semantic labels that can be used to train a segmentation model that then performs to high accuracy on real echo data, and 2) Release the generated datasets, as well as the code, for public usage.

The code is available at \link{https://github.com/david-stojanovski/echo\_from\_noise}. The generated diffusion model dataset, as described in Section \ref{sdm_model methods} is available at \link{https://zenodo.org/record/7921055\#.ZGYS\_9LMLmE}.

\section{Methods} \label{methods}
\begin{figure}[ht]
    \centering
    \makebox[\textwidth]{\includegraphics[width=\linewidth]{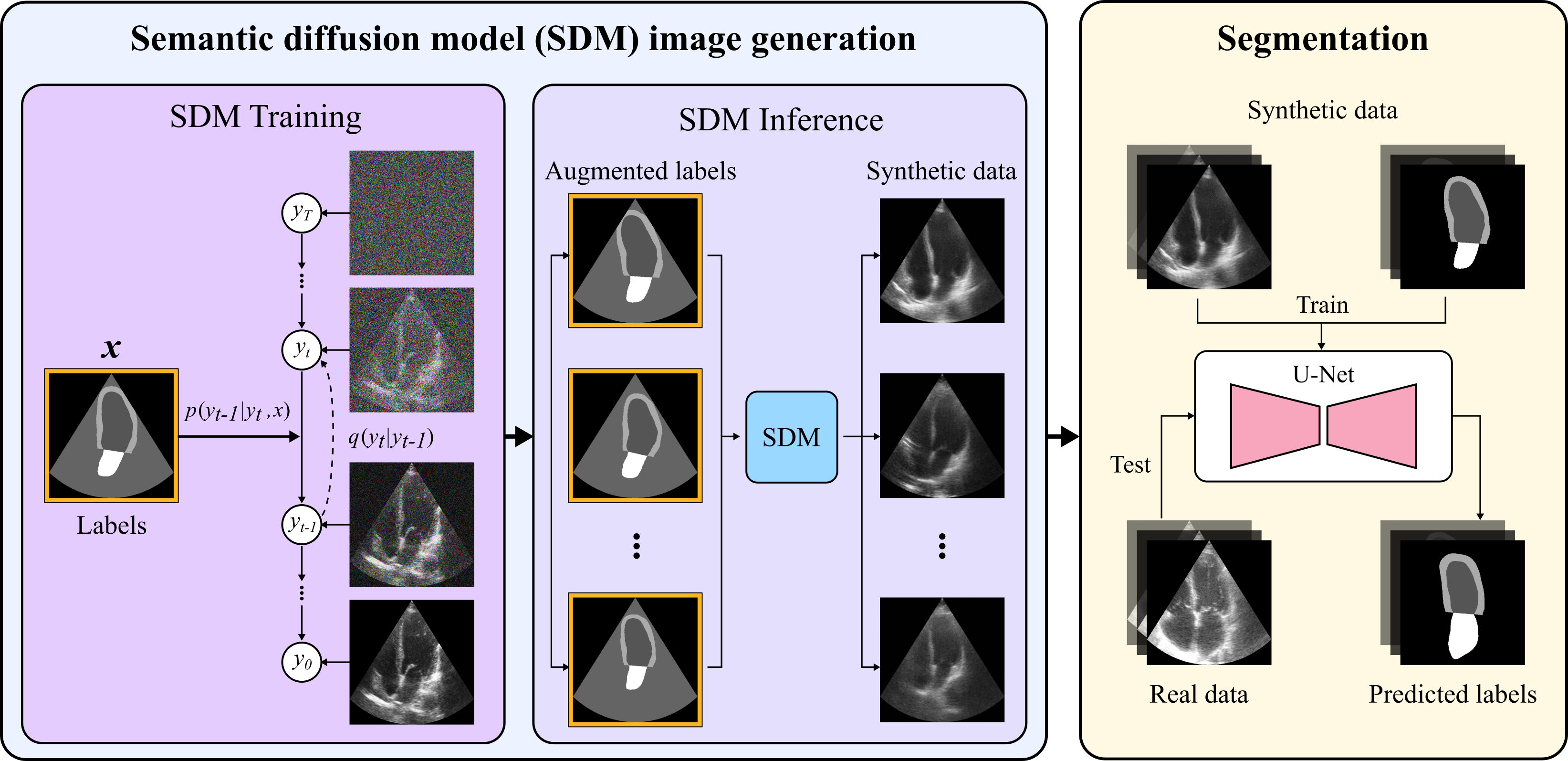}}

    \caption{Pipeline for generating synthetic ultrasound images from a semantic diffusion model (SDM), to be used in segmentation training and testing on real data. \textbf{SDM Training}: The SDM is trained to transform the noise to the realistic image through an iterative denoising process. \textbf{SDM inference}: The trained SDM is inferenced on augmented labels to generate corresponding realistic synthetic ultrasound images. \textbf{Segmentation}: The set of generated synthetic ultrasound images are used to train a segmentation network. The segmentation performance is tested on real data. \textit{x} denotes the semantic label map; $y_t$ denotes the noisy image at each time step $t$.}
    \label{pipeline_fig}
\end{figure}

As an overview, our image synthesis pipeline implements the Semantic Diffusion Model (SDM) proposed by Wang et al. \cite{Wang2022}, the details of which are described in Section \ref{sdm_model methods}. Subsequently, these generated images are used to train and validate a segmentation model. This model is then tested on an unseen dataset of real echocardiographic images, as described in Section \ref{segmentation_methods}. An overview of the pipeline is given in Fig. \ref{pipeline_fig}.

\subsection{Synthetic ultrasound generation: Semantic Diffusion Model} \label{sdm_model methods}
\subsubsection{Data:} \label{diffusion_model_dataset}
We used the CAMUS echocardiography dataset \cite{8649738} which contains 500 patients in total with semantically segmented left ventricle myocardium, endocardium, and left atrial surface for 4 chamber and 2 chamber images at both end-diastole (ED) and end-systole (ES) frames. The official test subset, of 50 patients, was reserved for testing segmentation networks. We used the remaining 450 patients for training and validating the generative models, by splitting them into 400 training and 50 validation. Note that only the ED frames were used for training and validating the two diffusion models (2C and 4C), totaling 400 training + 50 validation frames used to train each model.

An additional label was added to the label maps to describe the ultrasound sector (see Fig. \ref{pipeline_fig}), by applying a simple thresholding to the ultrasound images.

\subsubsection{Proposed model:}

We made a few notable modifications to the Semantic Diffusion Model (SDM), including some best practices proposed by Nichol et al. \cite{Nichol2021}.

Firstly we used a cosine noise schedule, instead of a linear schedule, to reduce the rate at which noise is added, thus increasing the contribution of the noise in later steps of the forward process. Secondly, the objective function being optimized is the summation of the predicted noise given an input label map at each time point and the Kullback–Leibler (KL) divergence between the estimated distribution and diffusion process posterior. We modified the weight of the KL divergence to $0.001$ to stabilize the optimization.

During inference, we removed the classifier-free guidance sampling for disentanglement, as it was found to add minimal perceptible difference to generated images. This allowed us to approximately halve the inference time. Briefly, the predicted noise of the model from a semantic label map is represented by:
\begin{equation}
    \hat{\epsilon}_{\theta}(y_t | x)  = \epsilon_{\theta}(y_t | x) + s \cdot (\epsilon_{\theta}(y_t | x) - \epsilon_{\theta}(y_t | \emptyset))
\end{equation}

\noindent where $\hat{\epsilon}_{\theta}(y_t | x)$ is the total-estimated noise in a ground truth image $y$ at time step $t$, given an input semantic label map $x$, $s$ is a user-defined guidance scale and $\epsilon_{\theta}(y_t | \emptyset)$ is estimated noise in a ground truth image given a null input, $\emptyset$. 

Two separate models were trained, a model on 4 chamber (4C) end diastolic frames and a model on 2 chamber (2C) end diastolic frames. Training and inference was performed using Pytorch 1.13 \cite{Paszke2019} on 8 $\times$ Nvidia A100 graphics processing units for 50,000 steps using an annealing learning rate and a batch size of 12. 

\subsection{Segmentation}
\subsubsection{Data:}
From the 400+50 CAMUS patient images used to build the diffusion models, we took the semantic maps (4 per patient: 2C and 4C both ED and ES) to produce four synthetic datasets: 2CED, 2CES, 4CED and 4CES. For each dataset, we took the 400+50 semantic maps and applied five random transformations (each being a combination of random affine and elastic deformation), to produce a total of 2000 (training) + 250 (validation) semantic maps per dataset. Affine transformation ranges for rotation degrees, translate, scale and shear were: $(-5, 5)^\circ$, $(0, 0.05)$, $(0.8, 1.05)$ and $5^\circ$ respectively. The number of control points and max displacement were $(10, 10, 4)$ and $(0, 30, 30)$. The corresponding echo images were generated by using the previously trained SDMs (the same SDM was used for ED and ES frames from a given view). A fifth dataset is built by aggregating the other four.

\begin{figure}[!h]
    \centering
    \makebox[\textwidth]{\includegraphics[width=0.8\linewidth]{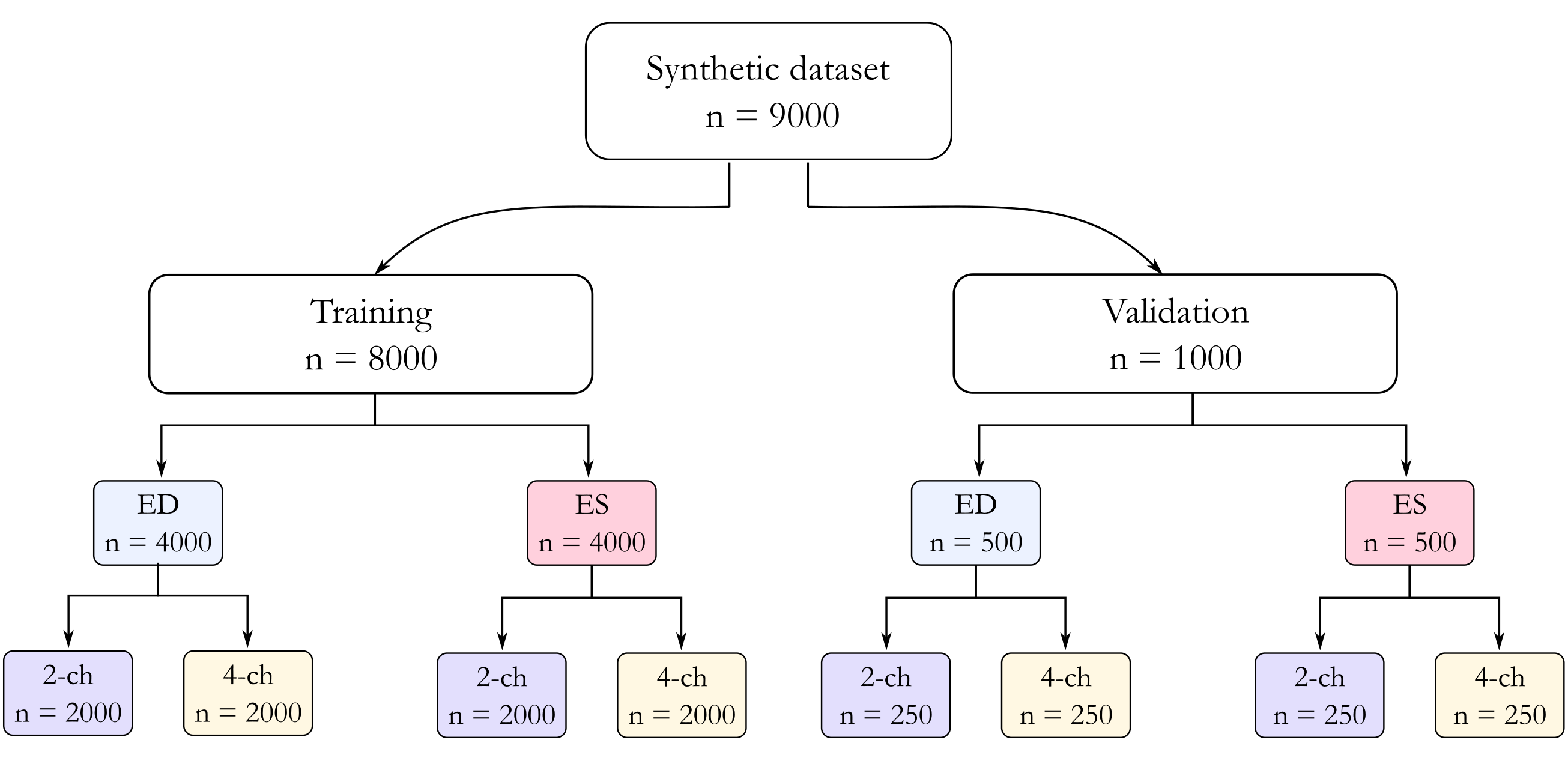}}
    \caption{Flow chart of semantic diffusion model dataset design. ED: end diastolic frames, ES: end systolic frames, 2-ch: echo 2 chamber images, 4-ch: echo 4 chamber images.}
    \label{dataset_fig}
\end{figure}

The exact generated datasets constituency is shown in Fig. \ref{dataset_fig}. The results given for the segmentation tasks are the values from testing on the official test split of the CAMUS dataset.

\subsubsection{Model:} \label{segmentation_methods}
We implemented an 8 layer U-Net model \cite{Ronneberger2015}, identical to the one in  \cite{Gilbert2021} for fair comparison. Five instances of this model were trained with the 5 aforementioned datasets. Segmentation accuracy was assessed using the 2D Dice score \cite{Thorvald1945}. The segmentation model was an in-house implementation of the standard U-Net using MONAI  \cite{the_monai_consortium_2020_4323059}, adapted for multiple labels and image resolution of (256, 256) as input. The network contains 8 layers (L) with $2^L$ channels in each layer and was trained for 300 epochs with the Adam optimizer and a learning rate of $1 \times 10^{-3}$, $\beta_1$ of 0.9 and $\beta_2$ of 0.999. These hyperparameters were chosen based on best practices proposed in literature \cite{Kingma2015}.

\subsection{Baseline CycleGAN model and data}

The CycleGAN network from \cite{Gilbert2021} was used as a benchmark for the SDM ultrasound image generation. The CycleGAN was trained with 2C and 4C slices from a public dataset of 1000 synthetic cardiac meshes \cite{Rodero2021}. Data was divided into a 70/15/15\% train/validation/test split and trained for 200 epochs. The training procedure was implemented as described in \cite{10.1007/978-3-031-16902-1_9}. The Dice score comparisons were made with the originally published results by Gilbert et al. \cite{Gilbert2021b}.

\section{Experiments and Results}
\begin{figure}[!htb]
    \centering
        \makebox[\textwidth]{\includegraphics[width=1\textwidth]{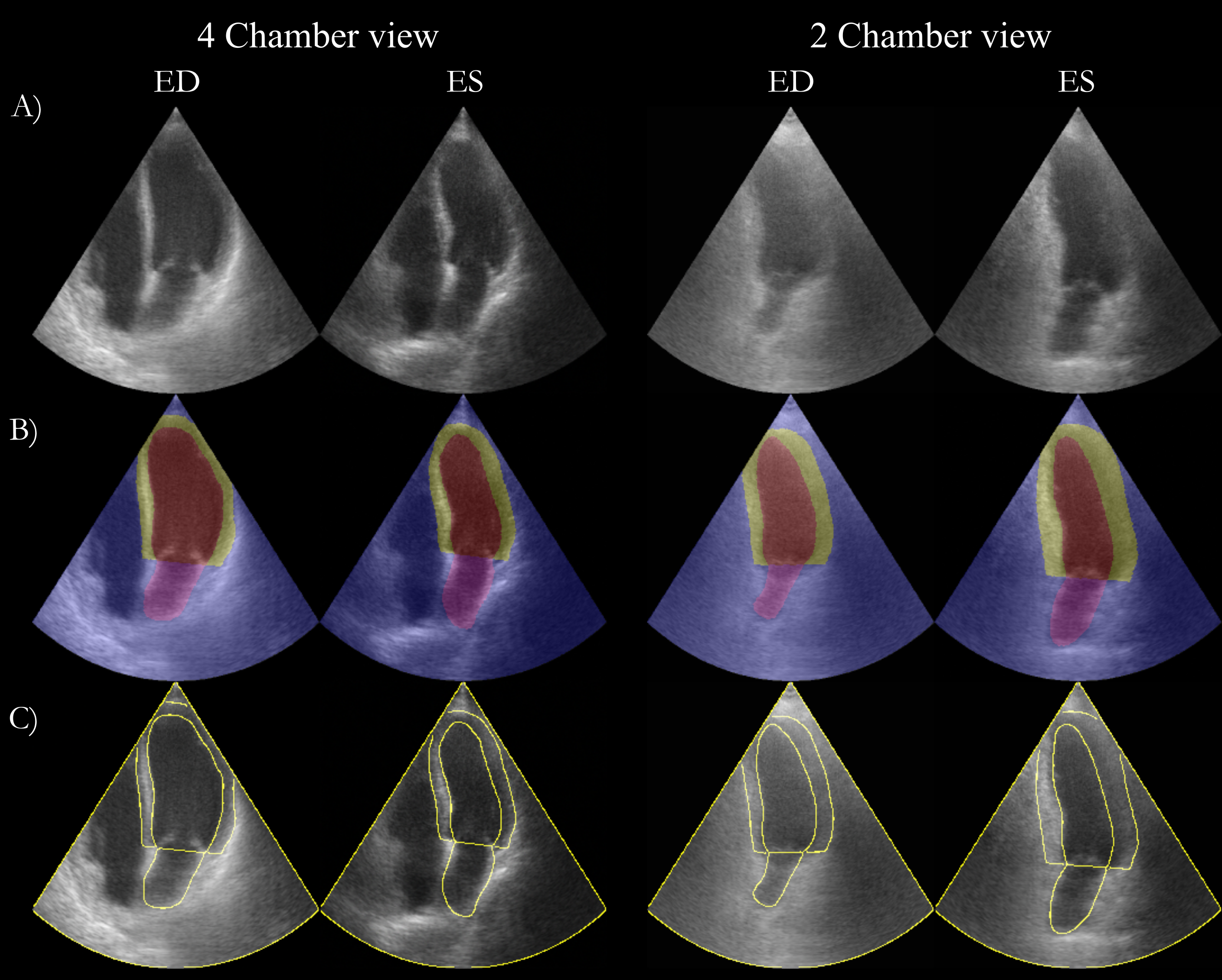}}
    \caption{Example images generated from our trained SDM networks. A) SDM synthetic images; B) SDM synthetic images overlaid with input semantic label map; C) Synthetic image with contour of input semantic label map. ED: End diastole; ES: End systole.}
    \label{example image fig}
\end{figure}

\begin{figure}[!htb]
    \centering
        \makebox[\textwidth]{\includegraphics[width=1.0\linewidth]{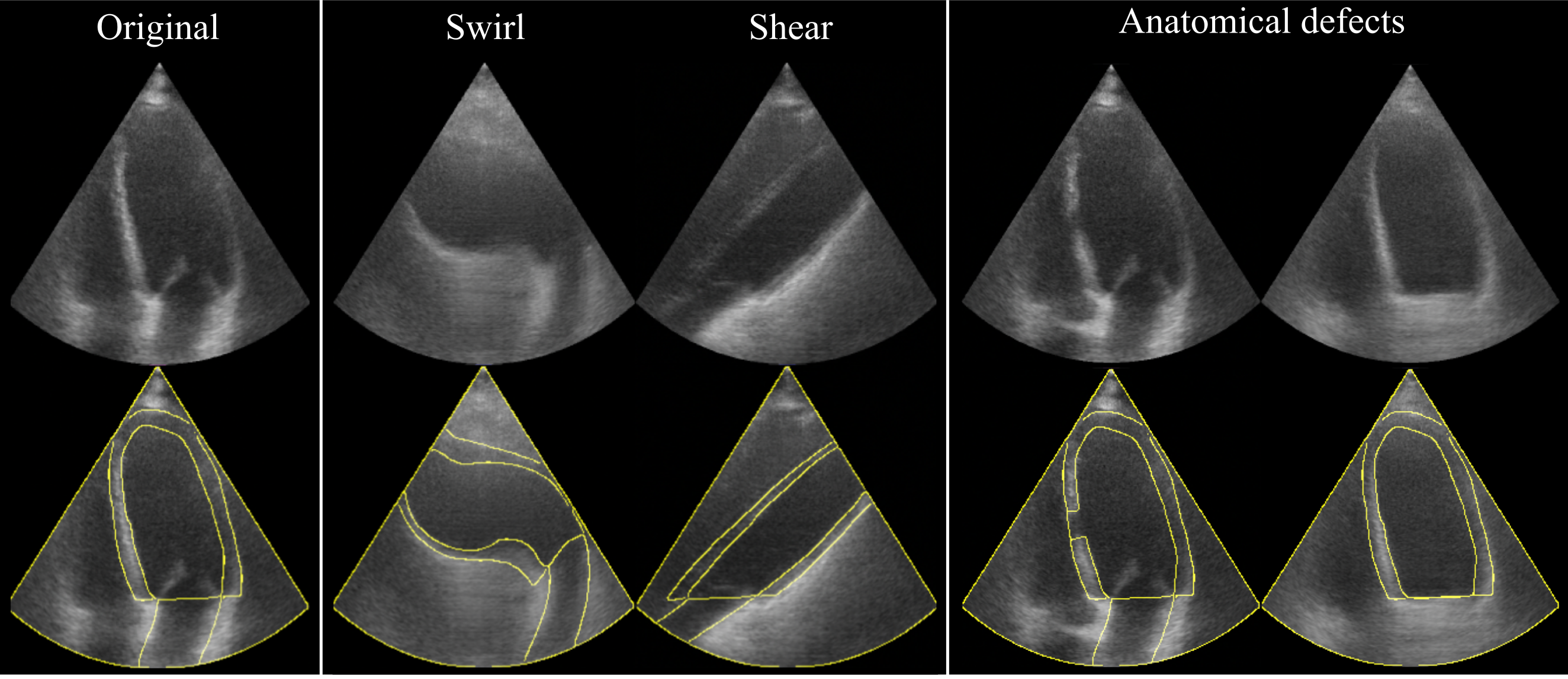}}

    \caption{Example images generated from extreme semantic map distortions. Top row: Synthetic images; Bottom row: Synthetic images with semantic label map contours. Anatomical defects: a hand-crafted septal defect (left), and missing left atrium (right).}
    \label{anatomical_aug_fig}
\end{figure}

\begin{figure}[!htb]
    \centering
    \makebox[\textwidth]{\includegraphics[width=1.0\linewidth]{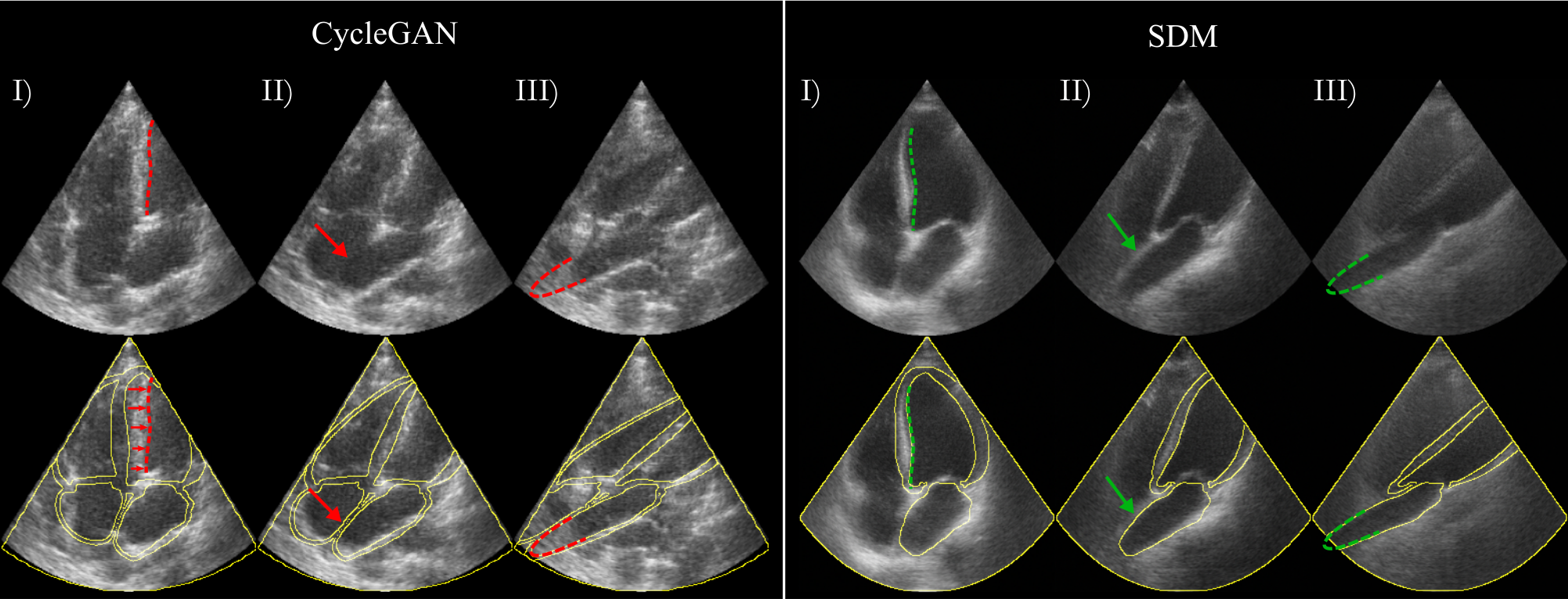}}
    \caption{Comparison of generated images from the SDM and our CycleGAN network. Top row: Generated synthetic images, bottom row: Overlay of the corresponding label map. Red dashed lines and arrows point to anatomical features missed or wrongly predicted by the CycleGAN model. Green dashed lines and arrows point to anatomical features correctly predicted by the SDM model.}
    \label{cyclegan_comparison}
\end{figure}

The final Dice scores on the unseen CAMUS test set of 50 patients are shown in Table \ref{dice_results}. Example images generated with the SDMs in all frames for a single patient are shown in Fig. \ref{example image fig}. We also present extreme augmentations in Fig. \ref{anatomical_aug_fig} to illustrate the robustness of the SDM model to inputs that would be out of the distribution of the training set. Fig. \ref{cyclegan_comparison} shows a comparison of our SDM model against a CycleGAN model for inferencing across a range of augmentations.

\setlength\tabcolsep{0pt}
\begin{table}[!h]\label{dice_results}
\centering
\caption{Dice scores for the CAMUS test set. $LV_{endo}$, $LV_{epi}$ and $LA$ denote left ventricular endocardium, epicardium and atrium respectively. Rows stating \textit{all frames} show mean and standard deviation for each label from a network trained on all views. \textit{All frames} refers to 2CH and 4CH images at both systole and diastole (4 images total).}
  \begin{tabular*}{\linewidth}{@{\extracolsep{\fill}} lcccc}
    \toprule
    \multirow{2}{*}{\textbf{Train/Validation Data}} &
      \multicolumn{4}{c}{\textbf{Dice Score (\%)}} \\
      & {$LV_{endo}$} & {$LV_{epi}$} & {\textit{LA}} & {\textit{Mean}} \\
      \midrule
    SDM 4 Chamber ED & 91.3$\pm 4.4$ & 94.0$\pm 3.2$ & 82.6$\pm 16.9$ & 89.3$\pm 4.9$ \\
    SDM 4 Chamber ES & 87.8$\pm 5.5$ & 92.7$\pm 2.7$ & 88.0$\pm 5.3$ & 89.5$\pm 2.3$ \\
    SDM 2 Chamber ED & 88.0$\pm 4.9$ & 89.3$\pm 5.5$ & 77.5$\pm 16.8$ & 85.0$\pm 5.3$\\
    SDM 2 Chamber ES & 87.0$\pm 5.7$ & 92.9$\pm 2.4$ & 87.9$\pm 7.3$& 89.3$\pm 2.6$\\
    \bottomrule
    \bottomrule
    All SDM frames & \textbf{88.6}$\pm 5.8$ & \textbf{91.9}$\pm 4.2$ & \textbf{85.2}$\pm 13.2$ & \textbf{88.6}$\pm 2.7$\\
    \textit{CycleGAN} \cite{Gilbert2021b} all frames & $79.4\pm9.9$ & $88.6\pm5.4$ & $71.3\pm23.7$ & 79.7$\pm7.1$\\
    All real CAMUS frames & $87.8\pm 6.8$ & $90.9\pm5.4$ & $82.6\pm14.4$ & 87.1$\pm3.4$\\ 
    \bottomrule
  \end{tabular*}
\end{table}

Using the CAMUS pretrained \textit{All SDM frames} model and performing testing on the EchoNet-Dynamic Atrial 4 Chamber at End Diastole dataset \cite{ouyang2019echonet}, the model achieved $87.83 \pm 7.21\%$ vs $85.6 \pm 7.0\%$ obtained by \textit{Gilbert et al.} when training a dataset-specific CycleGAN model.

Qualitatively, our results suggest that our SDMs are able to generate ultrasound images with superior overall image realism and propensity for adhering to anatomical input constraints, as well as the ability to generate images from extreme out-of-distribution semantic label maps. Representative examples of generated images are shown in Fig. \ref{example image fig}, \ref{anatomical_aug_fig} and \ref{cyclegan_comparison}. 

An example of our model vs CycleGAN (previous SOTA) is shown in  Fig. \ref{cyclegan_comparison}. Columns 2 and 3 represent examples of CycleGAN's limited ability to generate anatomically realistic images, even with a realistic anatomical guiding image. CycleGAN outputs show atrial walls merging, hallucinated septal wall thickening, or complete collapse of the ventricle. 

\section{Discussion and Conclusions}
Our anatomy-guided diffusion models generate realistic echo images that, compared to SOTA methods, adhere better to the anatomical constraints of a given label map, even when the prescribed anatomy is very different from the training set. Indeed, While CycleGAN can generate realistic ultrasound images, we observed that their ability to reproduce the anatomy under large deformation of the guide images is limited. Further, the anatomically accurate synthetic data generated with our model significantly improves the performance of the segmentation model on real images, showing the potential to address challenges involving rare medical conditions, data privacy, or limited data availability.

A segmentation network trained on the SDM-generated synthetic data significantly outperformed SOTA methods, and even training on the original real data, in segmentation of real 2 and 4 chamber ultrasound images. The LV endocardium, epicardium, and atrium were segmented with high accuracy (88.6$\pm 5.8$, 91.9$\pm 4.2$, 85.2$\pm 13.2$ \% Dice score respectively). These results represent an 11.6, 3.7 and 19.5\% improvement in relation to previous SOTA results. Moreover, our results showed reduced standard deviation for all labels, suggesting our model yields realistic images more consistently, leading to less variation in segmentation performance. These results also highlight the adaptability of SDMs for generating new data. 

Future work will include addressing variation across devices and clinical centers and investigating temporal aspects of synthetic data generation. The results presented in this paper show promise for synthetic data generation that can be used to train deep neural networks to high performance, addressing a crucial problem in medical imaging such as the availability of expert-labeled data.

\newpage
\bibliography{library}
\bibliographystyle{splncs04}

\end{document}